\documentclass[reprint,prd,nofootinbib,nobibnotes,amsmath,amssymb,aps,floatfix,superscriptaddress]{revtex4-2}
\usepackage{graphicx}
\usepackage{upgreek}
\usepackage{dcolumn}
\usepackage{bm}
\usepackage[colorlinks=true,linkcolor=red,urlcolor=blue,citecolor=blue]{hyperref}
\usepackage{float}
\begin{document}
\preprint{APS/123-QED}

\title{Calibration of the in-orbit center-of-mass of TaiJi-1}

\author{Xiaotong Wei}
\email{weixiaotong181@mails.ucas.ac.cn}
\affiliation{University of Chinese Academy of Sciences (UCAS), Beijing 100049, China}
\affiliation{International Centre for Theoretical Physics Asia-Pacific, UCAS, Beijing 100190, China}
\affiliation{Taiji Laboratory for Gravitational Wave Universe (Beijing/Hangzhou), UCAS, Beijing 100190, China}

\author{Li Huang}
\affiliation{University of Chinese Academy of Sciences (UCAS), Beijing 100049, China}
\affiliation{International Centre for Theoretical Physics Asia-Pacific, UCAS, Beijing 100190, China}
\affiliation{Taiji Laboratory for Gravitational Wave Universe (Beijing/Hangzhou), UCAS, Beijing 100190, China}

\author{Tingyang Shen}
\affiliation{University of Chinese Academy of Sciences (UCAS), Beijing 100049, China}
\affiliation{International Centre for Theoretical Physics Asia-Pacific, UCAS, Beijing 100190, China}
\affiliation{Taiji Laboratory for Gravitational Wave Universe (Beijing/Hangzhou), UCAS, Beijing 100190, China}

\author{Zhiming Cai}
\affiliation{Innovation Academy for Microsatellites, Chinese Academy of Sciences, Shanghai 201304, China}

\author{Jibo He}
\email{jibo.he@ucas.ac.cn}
\affiliation{University of Chinese Academy of Sciences (UCAS), Beijing 100049, China}
\affiliation{International Centre for Theoretical Physics Asia-Pacific, UCAS, Beijing 100190, China}
\affiliation{Taiji Laboratory for Gravitational Wave Universe (Beijing/Hangzhou), UCAS, Beijing 100190, China}
\affiliation{Hangzhou Institute for Advanced Study, UCAS, Hangzhou 310024, China}

\date{\today}

\begin{abstract}
Taiji program is a space mission aiming to detect gravitational waves in the low frequency band.
Taiji-1 is the first technology demonstration satellite of the Taiji Program in Space, with the gravitational reference sensor (GRS) serving as one of its key scientific payloads. 
For accurate accelerometer measurements, the test-mass center of the GRS must be positioned precisely at the center of gravity of the satellite to avoid measurement disturbances caused by angular acceleration and gradient. 
Due to installation and measurement errors, fuel consumption during in-flight phase, and other factors, the offset between the test-mass center and the center-of-mass (COM) of the satellite can be significant, degrading the measurement accuracy of the accelerometer. 
Therefore, the offset needs to be estimated and controlled within the required range by the center-of-mass adjustment mechanism during the satellite's lifetime. 
In this paper, we present a novel method, the Extended Kalman Filter combined with Rauch-Tung-Striebel Smoother, to estimate the offset, while utilizing the chi-square test to eliminate outliers. 
Additionally, the nonlinear Least Squares estimation algorithm is employed as a crosscheck to estimate the offset of COM. 
The two methods are shown to give consistent results, with the offset estimated to be $dx \approx $$-$$0.19$ mm, $dy \approx 0.64$ mm, and $dz \approx $$-$$0.82$ mm. 
The results indicate a significant improvement on the noise level of GRS after the COM calibration, which will be of great help for the future Taiji program. 
\end{abstract}
\maketitle

\section{\label{sec:lv1}Introduction}
In 2016, the LIGO collaboration made a groundbreaking discovery by detecting gravitational waves (GW)~\cite{PhysRevLett.116.061102}, thereby providing a direct verification of the predictions made by Albert Einstein in his general theory of relativity a century ago~\cite{einstein1916approximative}.
This discovery has had a profound impact on basic scientific research worldwide.
 
Space-based gravitational wave detection represents an intriguing frontier for future studies of the gravitational universe, as it can extend the reach of gravitational wave astronomy beyond that of the ground-based detectors, thereby allowing for a wider range of gravitational radiation sources to be observed~\cite{10.1093/nsr/nwx116}, which will provide invaluable information to deepen our understanding of the evolution of early universe and the nature of gravity. 
Several space-borne gravitational-wave observatories have been proposed, 
such as LISA~\cite{bender1998lisa,jennrich2011ngo,amaro2017laser}, DECIGO~\cite{PhysRevLett.87.221103}, ASTROD~\cite{Ni:2012eh}, Taiji~\cite{10.1093/nsr/nwx116,cyranoski2016chinese}, and Tianqin~\cite{luo2016tianqin}.

The Taiji program, initiated by the Chinese Academy of Sciences (CAS), is a space mission that aims to detect gravitational waves (GW) in the frequency band between 0.1~mHz and 1.0~Hz, which is important in the fields of astronomy and cosmology~\cite{cyranoski2016chinese,sathyaprakash2009physics,schutz1999gravitational}.
The Taiji program proposes to detect GW signals using the Michelson laser interferometer principle, where each end of the interferometer contains a test mass (TM) serving as a reference body. 
This reference body is required to be free from spurious accelerations relative to its local inertial frame, and any spurious accelerations will affect the detection of tidal deformations caused by gravitational waves. 

To facilitate the development of technology for the Taiji program, a three-step road map has been proposed~\cite{LUO2020102918,luo2020taiji}. 
As the first step, a technology demonstrator satellite, Taiji-1~\cite{luo2020taiji,Cai:2021xsz}, 
was launched on August 31, 2019. 
One of the key technologies validated by Taiji-1 is the Gravity Reference Sensor (GRS), which serves as an accelerometer and consists of sensors and electronic components~\cite{touboul1999accelerometers}. 
The sensor comprises an electrode housing and a TM surrounded by the sensing electrode, as shown in Fig.~\ref{fig:grs}. 

GRS has three axes, including one non-sensitive axis and two sensitive axes, the directions of $+\textrm{X}$, $+\textrm{Z}$, and $+\textrm{Y}$ in Fig.~\ref{fig:grs} correspond to the non-sensitive axis (radial direction), the first sensitive axis (flight direction), and the second sensitive axis, respectively.
The sensor utilizes capacitive sensing technology to measure the disturbance acceleration of Taiji-1. 
The resulting data is sent to the drag-free controller, which instructs the thruster to apply force to compensate for the disturbing force.
As the reference body of the future Taiji program interferometer, GRS needs to effectively mitigate the impact of all non-gravitational accelerations so that to reach the desired sensitivity of Taiji.
Moreover, the accuracy of GRS is crucial for the drag-free control system, as the GRS readout results are used as inputs for issuing commands to control the spacecraft.
GRS is susceptible to a range of noise sources, including Brownian noise arising from the surrounding air near TM, charge-induced noise due to TM charge accumulation, readout noise from voltage signals, temperature gradients, circuit noise, magnetic noise, self-gravity noise.

A precise positioning of the test-mass of the accelerometer at the center-of-mass (COM) of the Taiji-1 satellite is crucial to suppress non-gravitational and perturbed accelerations such as angular motion-related accelerations and accelerations due to gravity gradients~\cite{wang2010determination,rs14164030}. 
The COM of the accelerometer is adjusted 
to be at the COM of the satellite before launching. 
However, during the satellite's operation, the consumption of propellant causes the COM of the satellite to change relative to the satellite frame, leading to a shift of the accelerometer COM relative to the satellite COM over time. 
Therefore, it is crucial to regularly measure the deviation of the COM position of the two during the entire life cycle of the satellite, and use the COM adjustment mechanism to perform in-orbit adjustments so that the deviation is within a certain range~\cite{Armano:2018ucz,Dong2009,guzman2008subtraction}. 
 
In Section~\ref{sec:theory}, the principle of the COM calibration is presented. 
The accelerometer measurement model is described in Section~\ref{sec:model}. 
We discuss the use of the Extended Kalman filter model and Rauch-Tung-Striebel Smoother for COM calibration in Section~\ref{sec:EKF}, while the detection and removal of outliers principle is explained in Section~\ref{sec:outliers}. 
The performance of the COM calibration is evaluated in Section~\ref{sec:performance}, and the results and conclusions are summarized in Section~\ref{sec:Discussion and Conclusions}.

\section{Principle of COM calibration during in-orbit}
\label{sec:theory}
\begin{figure}[H]
\centering
\includegraphics[width=.4\textwidth]{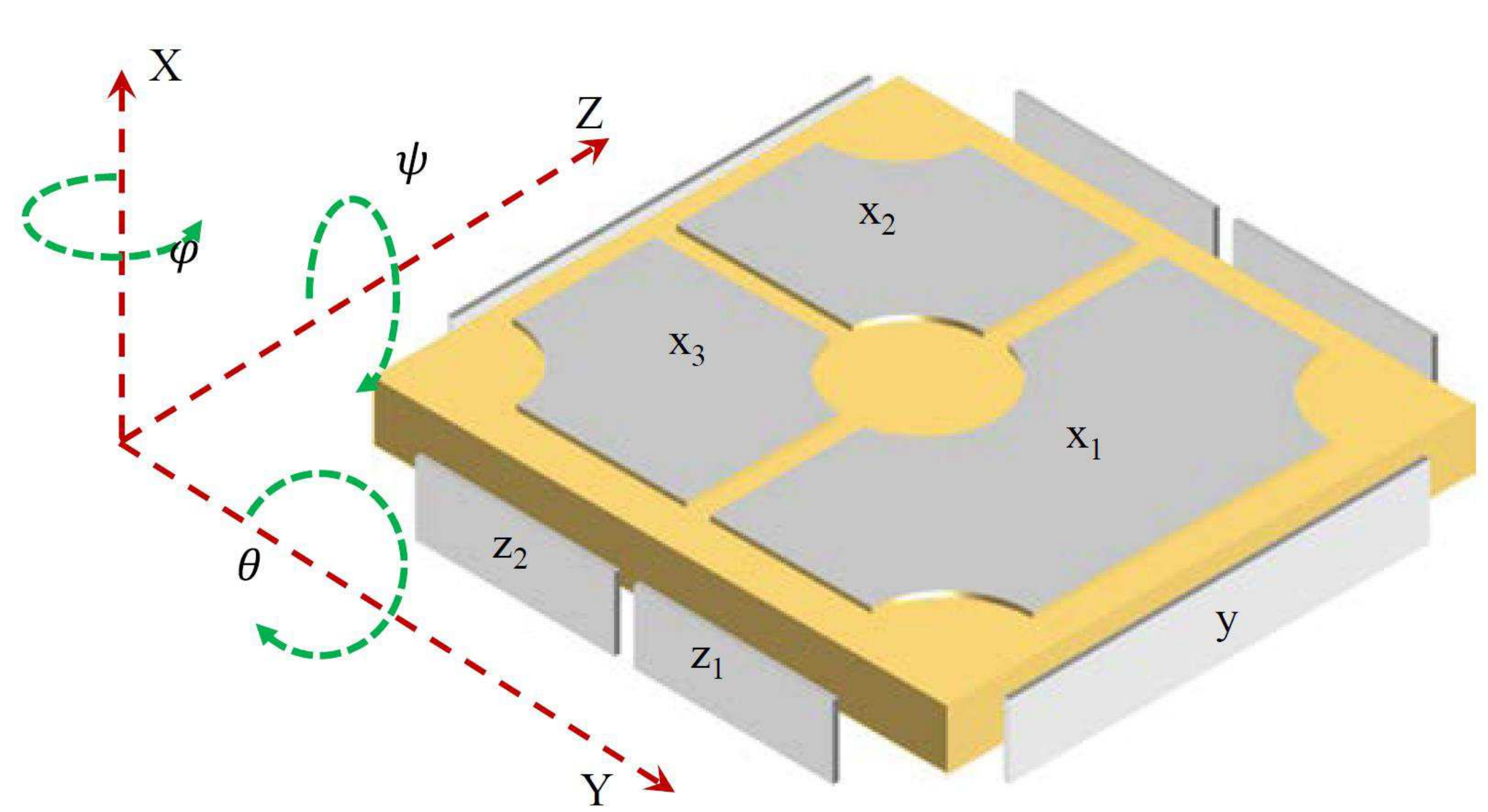}
\caption{The geometric structure of the core mechanical assembly of the GRS~\cite{peng2021system}.}
\label{fig:grs}
\end{figure}

As shown in  Fig.~\ref{fig:grs}, the primary components of the high-precision electrostatic levitation accelerometer include a free fall TM, a set of capacitive electrode plate that surrounds the TM (together forming a sensitive probe), and a peripheral capacitive sensing and electrostatic feedback control circuit. 
This circuit enables the detection of position and attitude changes between the TM and the electrodes, as well as the measurement of acceleration through its feedback voltage.
There are six sensing, control, and feedback circuits that use the same principle to measure three translational accelerations and three angular accelerations of the TM concurrently. 

An offset of the COM of the electrostatic accelerometer from the COM of the satellite can cause measurement disturbances, primarily attributed to the angular motion of the satellite. 
Therefore, an estimate of the offset can be achieved using appropriate algorithms based on the relationship between the angular motion and the measurement disturbance. 

\begin{figure}[H]
\centering
\includegraphics[width=.45\textwidth]{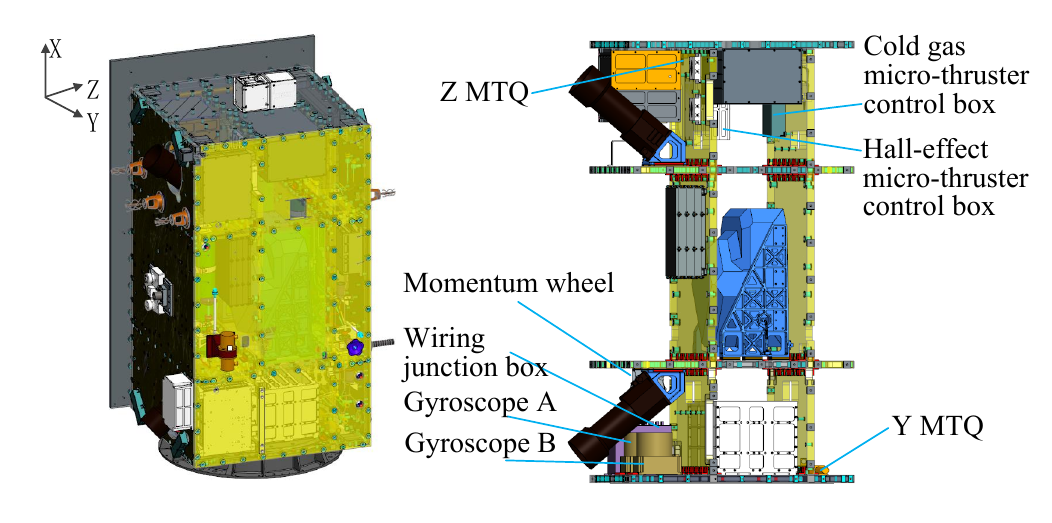}
\caption{The layout of the Taiji-1 attitude control system (MTQ stands for magnetic torquer).}
\label{fig:taiji1}
\end{figure}

The on-orbit center-of-mass calibration experiment is conducted using the Taiji-1 attitude control system, which consists of the attitude sensors and the actuators, as shown in Fig.~\ref{fig:taiji1}. 
The additude sensors includes star sensors, three-axis magnetometers, and gyroscopes.
The actuators include cold-gas micro-thrusters, Hall-effect micro-thrusters, magnetic torquers, and momentum wheel.
The X-axis magnetic torquer is mounted on the $+\textrm Z$ side panel of the satellite, the Y-axis magnetic torquer is mounted on the bottom panel, and the Z-axis magnetic torquer is mounted on the top panel. 
The direction of the positive magnetic moment aligns with the satellite's three axes, as shown in the left figure of Fig.~\ref{fig:taiji1}.

In the calibration experiment, periodic torques, generated by the magnetic torquers, is applied to the satellite, inducing a cyclic oscillation by the Earth's magnetic field, which will cause disturbance to the accelerometer measurements and ensure its magnitude is sufficiently large to disregard other disturbance effect, such as solar pressure torque and aerodynamic torque. 
Then, the offset can be modeled from readout of accelerometer and star tracker. 
Moreover, the accuracy of GRS is crucial for the drag-free control system, as the GRS readout results are used as inputs for issuing commands to control the spacecraft.

The TM is made of titanium alloy TC4, with a magnetic susceptibility of $3.2 \times 10^{-6}$, therefore the effects of the coupling between the TM and the magnetic torquers is much smaller than the present GRS accuracy and can be neglected.

\section{Accelerometer measurement Model}
\label{sec:model}
The accelerometer measurement output model of the relative acceleration of TM and the electrode cage for Taiji-1 is presented as follows~\cite{Cai:2021xsz, wang2010determination, rs14164030}, 
\begin{equation}
a_{\textrm{out}}=\ddot{d}+\dot{\omega} \times d+2 \omega \times \dot{d}+\omega \times(\omega \times d)+a_\textrm{g}+ a_\textrm{ng}
\label{eq:linear}
\end{equation}
where $a_\textrm{out}$ represents the theoretical measurements of the accelerometer, $d$ represent the COM offset between the accelerometer and the satellite, $\dot{d}$ and $\ddot{d}$ denote the first and second partial derivation of $d$ relative to time. 
$\omega$ and $\dot{\omega}$ are the angular velocity and the angular acceleration respectively. 
$a_\textrm{g}$ is the acceleration due to the gravitational gradient, the accelerometer measurement model ignores the perturbations due to the solid Earth tides, ocean tides, rotational deformations, the planets including the Sun and Moon, and general relativity which can be neglected as for the calibration span is short enough. 
Furthermore, $a_\textrm{ng}$ represents the non-gravitational acceleration on the satellite, such as atmospheric drag, solar radiation pressure, and the Earth radiation pressure. 

During the offset calibration period, the deviation of the COM of the TM from the COM of the satellite can be approximated as a constant offset, given the short measurement time. 
Therefore, $a_\textrm{out}$ can be expressed as,
\begin{equation}
a_{\textrm{out}}=\dot{\omega} \times d+\omega \times(\omega \times d)+a_\textrm{g}+ a_\textrm{ng}.
\label{eq:linearend}
\end{equation}
Considering the relatively smooth change of acceleration caused by non-conservative forces and gravity gradient, it can be approximated as a linear change within a limited time span. 
Therefore, the calibration interval is designed to be several minutes. 

The attitude control system of the Taiji-1 satellite is equipped with a star tracker and a gyroscope to respectively measure the attitude angle of the satellite relative to the inertial coordinate system and the angular velocity $\omega$. 
Additionally, the angular accelerations $\dot{\omega}$ are obtained by a second-order polynomial fitting method. 

Assuming negligible scale factors and misalignment errors, after substituting the $\omega$ and $\dot{\omega}$, the accelerometer measurement model can be expressed as follows,
\begin{equation}
A_\textrm{out}=\widetilde{A} d+\alpha t+\beta+A_\textrm{n},
\label{eq:A_out}
\end{equation}
where $\tilde{A}$ can be expressed as
\begin{equation}
\tilde{A}=\left[\begin{array}{ccc}
-\omega_\textrm{y}^{2}-\omega_\textrm{z}^{2} & \omega_\textrm{x} \omega_\textrm{y}-\dot{\omega}_\textrm{z} & \omega_\textrm{x} \omega_\textrm{z}+\dot{\omega}_\textrm{y} \\
\omega_\textrm{x} \omega_\textrm{y}+\dot{\omega}_\textrm{z} & -\omega_\textrm{x}^{2}-\omega_\textrm{z}^{2} & \omega_\textrm{z} \omega_\textrm{y}-\dot{\omega}_\textrm{x} \\
\omega_\textrm{x} \omega_\textrm{z}-\dot{\omega}_\textrm{y} & \omega_\textrm{z} \omega_\textrm{y}+\dot{\omega}_\textrm{x} & -\omega_\textrm{y}^{2}-\omega_\textrm{x}^{2}
\end{array}\right].
\label{eq:A_tilde}
\end{equation}
The corresponding axis of the spacecraft can be represented by $\omega_\textrm{i}$ and $\dot{\omega}_\textrm{i}$ ($\textrm{i}=\textrm{x},\textrm{y},\textrm{z}$) for angular acceleration and angular velocity, respectively. 
The linear slope is represented by $\alpha$ and the constant bias by $\beta$. 
In this study, we removed the linear effect by detrending $A_\textrm{out}$. Therefore, the model of $A_\textrm{out}$ utilized in this study is,
\begin{equation}
A_{\rm out}=\widetilde{A} d + A_\textrm{n}
\label{eq:used_A_out}
\end{equation}
where $A_\textrm{n}$ is the measurement noise.

\section{Extended Kalman  filter model and Rauch-Tung-Striebel Smoother for COM calibration}
\label{sec:EKF}
Kalman filter is a high-efficiency recursive filter~\cite{10.1115/1.3662552}. 
The filtering theory proposed by Kalman is only applicable to linear systems. 
An Extended Kalman Filter (EKF) 
was proposed in Ref.~\cite{10.1115/1.3425006,bucy1971digital}, and can be applied to the nonlinear field. 

The accelerometer measurement model used in this study is given by Eq.~\ref{eq:used_A_out}, and one can define the state variable as the COM offset and use the Kalman filter to estimate the offset. 
Here, the state vector is denoted as $X = [d_\textrm{x}, d_\textrm{y}, d_\textrm{z}]$ with $\dot{d_\textrm{i}} = 0$ for $\textrm{i}=\textrm{x},\textrm{y},\textrm{z}$. 
The state equation, as derived in Ref.~\cite{Dong2009}, is,
\begin{equation}
\hat{X}_{k}=\Phi_{k,k-1} X_{k-1}.
\end{equation}
Here, $k$ signify the step of filter, $\Phi_{k,k-1}$ represents the state transition matrix from step $k-1$ to step $k$. 
As no manipulation is performed on the state vector, the state transition matrix can be written as $\Phi_{k,k-1} = I$, where $I$ is the identity matrix. 

The output of the accelerometer can be defined as the observation equation,
\begin{equation}
Z_\textrm{out,\textit{k}}=H_{k} {X}_{k}+V_{k},
\end{equation}
where $H_{k}=\tilde{A}_{k}$, and $V_{k}$ is the discrete measurement noise that satisfies,
\begin{equation}
E\left\{V_{k}\right\}=0, Cov\left\{V_{k}\right\}=R_{k}.
\end{equation}
Here, $R_{k}$ denotes the variance matrix of the measurement noise,
assuming that all $V_{k}$ are independent, unbiased, and possess finite variance which implies that $R_{k}$ is a diagonal matrix. 

The predicted observation equation is given by, 
\begin{equation}
\hat{Z}_\textrm{out,\textit{k}}=H_{k} \hat{X}_{k}.
\end{equation}
The estimated covariance matrix is given by,
\begin{equation}
\hat{P}_{k}=\boldsymbol{\Phi}_{k,k-1} P_{k-1} \boldsymbol{\Phi}_{k,k-1}^{\mathrm{T}}+Q_{k-1},
\end{equation}
where $Q$ denotes the system noise variance matrix, which is assumed to be zero in this study. 

The Kalman gain is
\begin{equation}
K_{k}=\hat{P}_{k} H_{k}^{\mathrm{T}}\left(H_{k} \hat{P}_{k} H_{k}^{\textrm{T}}+R_{k}\right)^\mathrm{-1},
\end{equation}
and the state update value after Kalman filter is,
\begin{equation}
X_{k}=\hat{X}_{k}+K_{k}\left(Z_\textrm{out,\textit{k}} - \hat{Z}_\textrm{out,\textit{k}}\right).
\end{equation}
Here, $\hat{X}_{k}$ represents the prior estimate value, and $X_{k}$ is the posterior estimate. 

The update of the error covariance matrix is,
\begin{equation}
P_{k}=\left(I-K_{k} H_{k}\right) \hat{P}_{k}\left(I-K_{k} H_{k}\right)^{\mathrm{T}}+K_{k} R_{k} K_{k}^{\mathrm{T}}.
\end{equation}
After the Kalman filter is applied, the difference between the predicted and observed measurements is known as the filtered residual. 
The filtered residual can be used to evaluate the accuracy of the filter. 
The covariance matrix of the filtered residual provides information on the uncertainty of the estimation. 
Therefore, it is important to analyze both the filtered residuals $r_{k}$ and the covariance matrix of the filtered residuals $R_{k}^\textrm{K}$, as follows,  
\begin{equation}
r_{k}=Z_\textrm{out,\textit{k}}-H_{k} X_{k},
\end{equation}
\begin{equation}
R_{k}^\textrm{K}=R_{k}-H_{k} P_{k} H_{k}^\mathrm{T}.
\end{equation}
The characteristic property of the Kalman filter is that the filtered residual vectors are uncorrelated, and even independent in the case of Gaussian distribution. 

After applying the aforementioned Kalman Filter, the state estimator of the dynamic system can be further refined using the Rauch-Tung-Striebel (RTS) smoother, as described in Ref.~\cite{ruach1965maximum}. 
The smoothing equations are given by the following formulas,
\begin{equation}
\hat{X}_{k+1}=\Phi_{k+1,k} X_{k},
\end{equation}
\begin{equation}
\hat{P}_{k+1}=\Phi_{k+1,k} P_{k} \Phi_{k+1,k}^\mathrm{T}+Q_{k},
\end{equation}
\begin{equation}
G_{k}=P_{k} \Phi_{k+1,k}^\mathrm{T} \hat{P}_{k+1}^\mathrm{-1}, 
\end{equation}
\begin{equation}
X_{k}^\textrm{S}=X_{k}+G_{k}\left(X_{k+1}^\textrm{S}-\hat{X}_{k+1}\right), 
\end{equation}
\begin{equation}
P_{k}^\textrm{S}=P_{k}+G_{k}\left(P_{k+1}^\textrm{S}-\hat{P}_{k+1}\right) G_{k}^\mathrm{T},
\end{equation}
\begin{equation}
r_{k}^\textrm{S}=Z_\textrm{out,\textit{k}}-H_{k} X_{k}^\textrm{S},
\end{equation}
\begin{equation}
R_{k}^\textrm{S}=R_{k}-H_{k} P_{k}^\textrm{S} H_{k}^\mathrm{T}.
\end{equation}

The RTS smoother provides smoothed estimates of the state mean and state covariance at time step $k$, denoted as $X_{k}^\textrm{S}$ and $P_{k}^\textrm{S}$, respectively. 
The smoother gain on time step $k$, denoted as $G_{k}$, corrects the RTS smoother estimate. 
The recursion is initialized at the last time step $\textrm{T}$ of Kalman filter with $X_\textrm{T}^\textrm{S} = X_\textrm{T}$ and $P_\textrm{T}^\textrm{S} = P_\textrm{T}$. 
The smoothed residuals are denoted as $r_{k}^\textrm{S}$, and the covariance matrix of smoothed residuals is denoted as $R_{k}^\textrm{S}$. 

To obtain a more accurate result, a combined method based on Kalman filter and the RTS Smoother (KF-RTS) and outliers removal is proposed. 
The data is filtered with the Kalman filter, and smoothed by the RTS smoother, which takes into account all the available valid data points. 
A chi-square confidence test is performed during the smoothing process for assessing the data quality and removing outliers. This KF-RTS process is iterated until there are no outliers. 

A possible drawback of the filter algorithm is the fact that one needs an initial value of the state vector together with its covariance matrix. 
This can be obtained by fitting a small number of measurements at the start of the track by a conventional least-squares fit, but this is not an elegant solution. 
The other possibility is to start with an arbitrary state vector and an infinite covariance matrix, i.e. a large multiple of the identity matrix. 
This is completely in the spirit of the filtering approach, but may lead to numerical instabilities in the computation of the gain matrix, since the infinities have to cancel in order to give a finite gain matrix. 
This may be difficult on a computer with a short word length. 

In this article the initial value of the state vector is set to zero and its covariance is chosen to be 0.001, which is large enough. 
The measurement noise is calculated by selecting a segment of stationary data from the corresponding data. 

Here, we use the nonlinear least squares (NLLS) method~\cite{doi:10.1057/jors.1985.68} for parameter estimation, as a crosscheck of the KF-RTS method. 
As is well-known, the key challenge for NLLS is to find the value of $\hat{\theta}$ that minimizes the function $F(\theta)$
\begin{equation}
F(\mathbf{\theta}) \equiv \frac{1}{2}\sum_{i=1}^{m}\left(f_{i}(\mathbf{\theta})\right)^\mathrm{2}
\end{equation}
where $f_{i}(\theta) \equiv y_{i}-\mathrm{Model}(\theta,\mathrm{input})$. 
The Levenberg-Marquardt (LM) algorithm~\cite{levenberg1944method,10.2307/2098941}, which is an efficient method to minimize $F(\theta)$, is used to find the optimal parameters in this article.

\section{Detection and removal of outliers}
\label{sec:outliers}

During the Kalman filtering and RTS smoothing, each data point was utilized to obtain an optimal estimate of the state value, and this process also allows for an assessment of the quality of each data point. 

The chi-square value~\cite{cochran1952chi2} is commonly used for assessing the quality of data and detecting  outliers, which can be caused by spacecraft maneuvering, non-Gaussian noise, electronic noise, or other coupled noise, and deviate significantly from the normal sequence of measurements. 

The residuals of the global fit can be utilized to identify measurements with large residuals as potential outliers. 
Kalman filters and smoother enable the exploitation of complete information locally, to determine the validity of a measurement with high probability. 
The smoothed residual chi-square value can serve as a useful decision criterion for the data quality of the measurement,
\begin{equation}
\chi^2=(r_{k}^\textrm{S})^\mathrm{T} (R_{k}^\textrm{S})^\mathrm{-1} r_{k}^\textrm{S}.
\end{equation}
It is demonstrated that the test on smoothed residual chi-square consistently outperforms the test on filtered residual chi-square~\cite{Fruhwirth:1987fm}. 
Thus, searching for possible outliers should be carried out during smoothing, as it allows the utilization of complete parameter information. 

During the smoothing process, the measurement point $Z_\textrm{out,\textit{k}}$ can be removed from the smoothed estimate $X_{k}^\textrm{S}$ to obtain ${X_{k}^\textrm{S}}^{*}$, which represents the optimal estimate of the system state at step $k$ using all data information except $Z_\textrm{out,\textit{k}}$. 
This optimal estimate can be utilized for the detection and removal of outliers. To remove $Z_\textrm{out,\textit{k}}$ from the estimate $X_{k}^\textrm{S}$, an inverse Kalman filter can be applied with the covariance matrix of $Z_\textrm{out,\textit{k}}$ taken as negative. 
This step of the filter is described in Ref.~\cite{Fruhwirth:1987fm}, and 
the smoothed estimate of $X_{k}$ without using $Z_\textrm{out,\textit{k}}$, ${X_{k}^\textrm{S}}^{*}$, can be calculated as, 
\begin{equation}
{X_{k}^\textrm{S}}^{*}={X}_{k}^\textrm{S}+{K_{k}^\textrm{S}}^{*}\left(A_\textrm{out,\textit{k}}-{H}_{k} {X}_{k}^\textrm{S}\right),
\end{equation}
in which 
\begin{equation}
{K_{k}^\textrm{S}}^{*}={P}_{k}^\textrm{S} {H_{k}}^\mathrm{T}\left({H}_{k} {P}_{k}^\textrm{S} {H_{k}}^\mathrm{T} - {R}_{k}\right)^\mathrm{-1},
\end{equation}
and 
\begin{equation}
{P_{k}^\textrm{S}}^{*}=\left({I}-{K_{k}^\textrm{S}}^{*} {H}_{k}\right) {P}_{k}^\textrm{S}. 
\end{equation}

If $Z_\textrm{out,\textit{k}}$ is a valid measurement and the covariance matrix of its Gaussian readout error is known, the quantity $\chi^2$ follows a chi-square distribution with $N_\textrm{z}$ degrees of freedom, where $N_\textrm{z}$ is the dimension of $Z_\textrm{out,\textit{k}}$. 

The measurement can be identified as an outlier if the value of $\chi^2$ exceeds a certain threshold $c$. 
This threshold is chosen as the $(1-\gamma)$ quantile of the corresponding $\chi^2$ distribution, where $\gamma$ represents the probability of rejecting a valid measurement and is chosen to be $\gamma=0.001$ in this paper. Other choices of $\gamma$ value, 0.005, 0.01, 0.05 and 0.1 have been tried, and the effects on the estimated COM offset is found to be negligible. 

The measurement $Z_\textrm{out,\textit{k}}$ can be removed permanently from the list as an outlier, and the RTS smoother can continue with ${X_{k}^\textrm{S}}^{*}$ and ${P_{k}^\textrm{S}}^{*}$ instead of ${X_{k}^\textrm{S}}$ and ${P_{k}^\textrm{S}}$ for updating the estimates $X_{j}^\textrm{S}$ when $j>k$. 
To remove all outliers, this Kalman filter and RTS smoother must be recomputed without the outliers and iterated until convergence is achieved.

\section{Results of COM calibration}
\label{sec:performance}
\begin{figure}
\centering
\includegraphics[width=.48\textwidth]{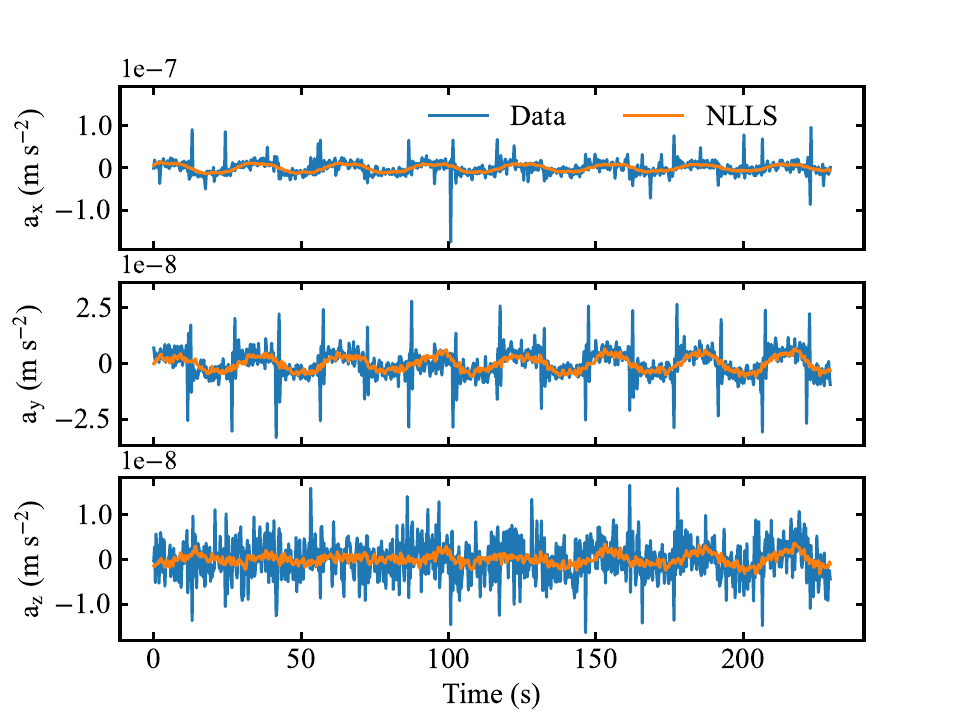}
\caption{The calibration experiment data and the NLLS fit results (with outliers).}
\label{fig:NLLS_with}
\end{figure}

\begin{figure}
\centering
\includegraphics[width=.48\textwidth]{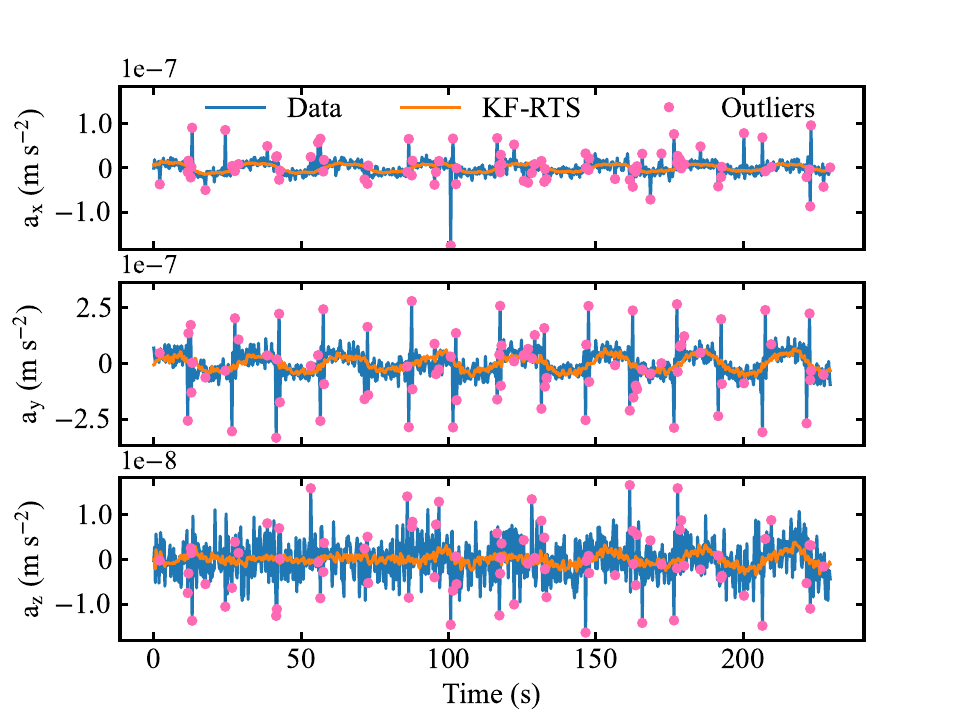}
\caption{The calibration experiment data, together with the first round of outliers detected through the $\chi^2$ test and KF-RTS smoother results obtained before removing the outliers.}
\label{fig:Outliers}
\end{figure}

\begin{figure}
\centering
\includegraphics[width=.48\textwidth]{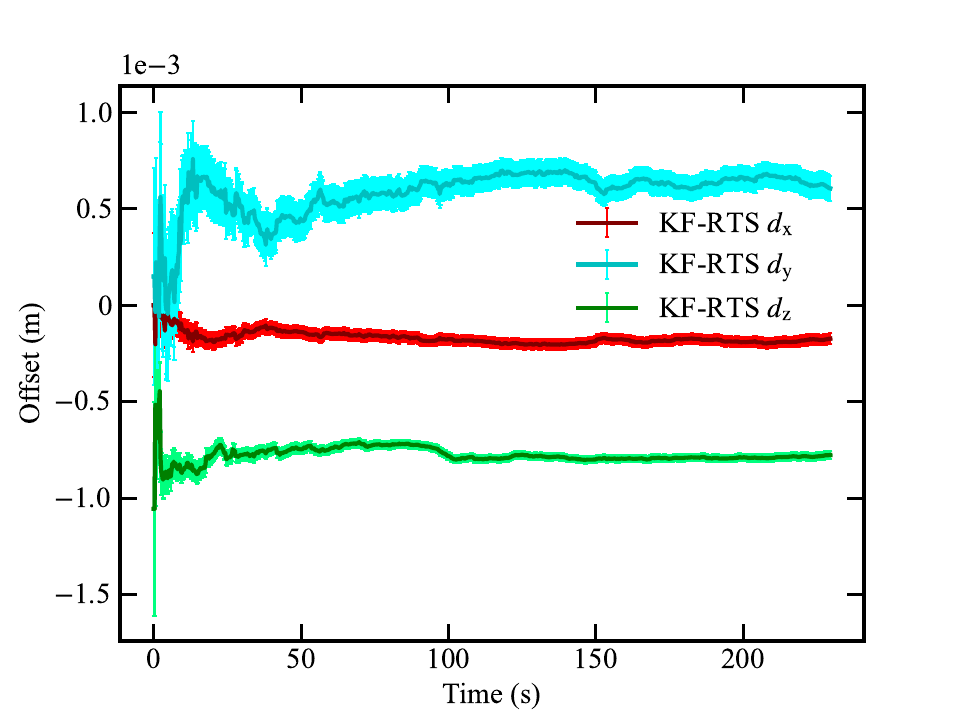}
\caption{The first KF-RTS smoother result of $d_\textrm{x}$ $d_\textrm{y}$ $d_\textrm{z}$.}
\label{fig:KF-RTS_dxdydz}
\end{figure}

The COM bias estimation in this paper is performed using the two algorithms discussed earlier. 
Figure~\ref{fig:NLLS_with} presents a comparison between the linear acceleration results obtained from the NLLS estimation and the original data, while Fig.~\ref{fig:Outliers} illustrates the comparison between the linear acceleration results obtained from the extended Kalman filter algorithm and the original data. 
Both methods exhibit excellent agreement with the original data. 
The results of COM calibration using the KF-RTS algorithm before removing outliers are shown in Fig.~\ref{fig:KF-RTS_dxdydz}, with the shaded area indicating the range of one standard deviation. 

\begin{figure}
\centering
\includegraphics[width=.48\textwidth]{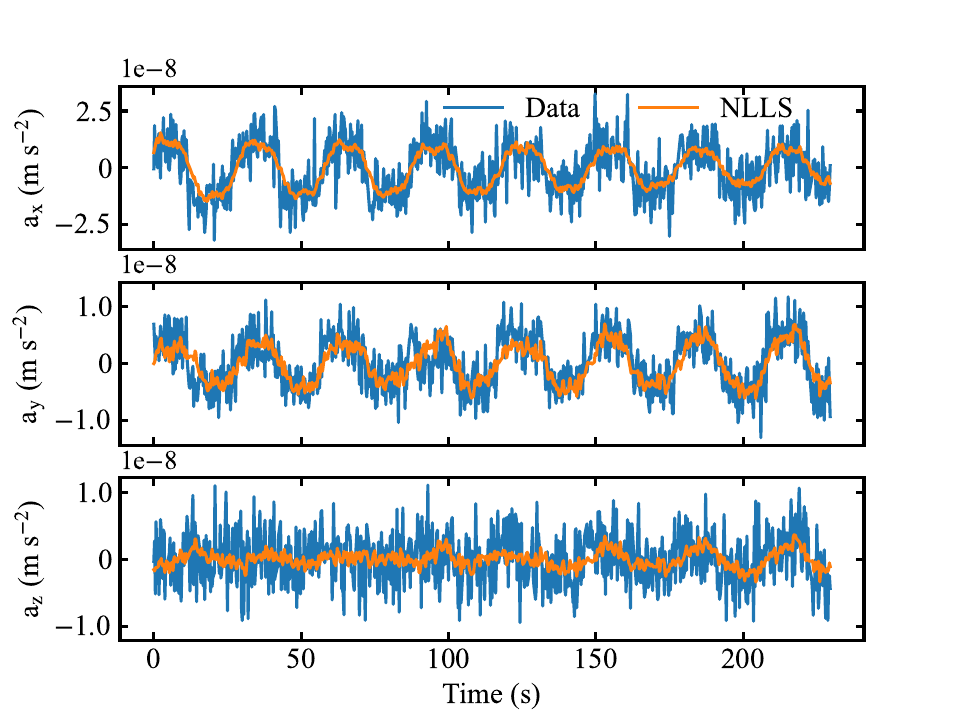}
\caption{The calibration experiment data and the NLLS fit results obtained after removing outliers.}
\label{fig:NLLS_without_outliers}
\end{figure}

\begin{figure}
\centering
\includegraphics[width=.48\textwidth]{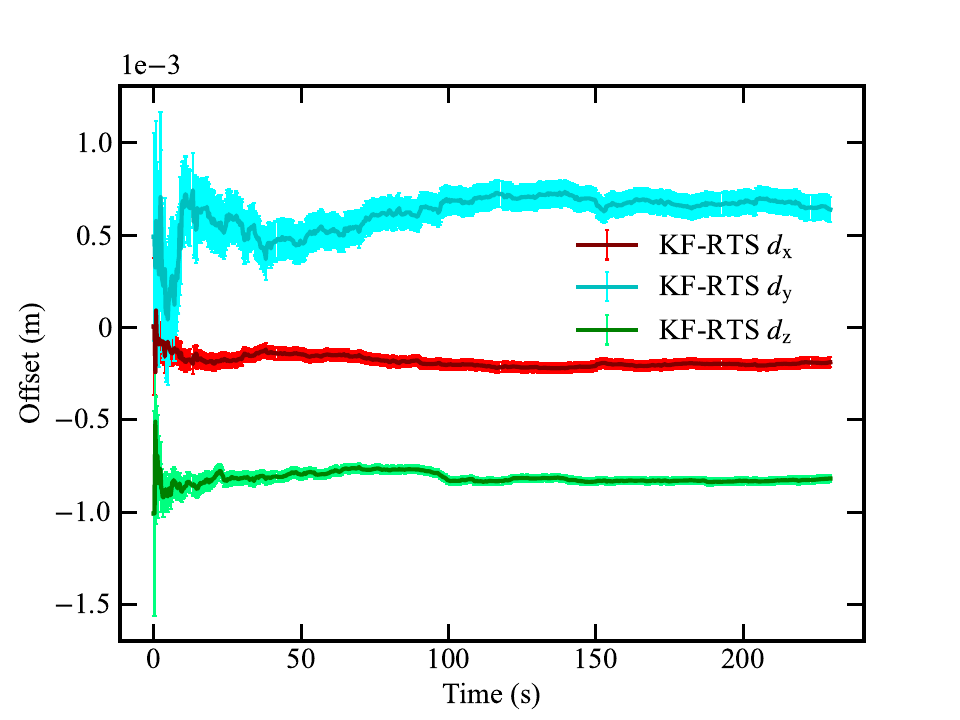}
\caption{The final KF-RTS smoother results of $d_\textrm{x}$ $d_\textrm{y}$ $d_\textrm{z}$.}
\label{fig:Final_KF-RTS_dxdydz}
\end{figure}

\begin{figure}
\centering
\includegraphics[width=.48\textwidth]{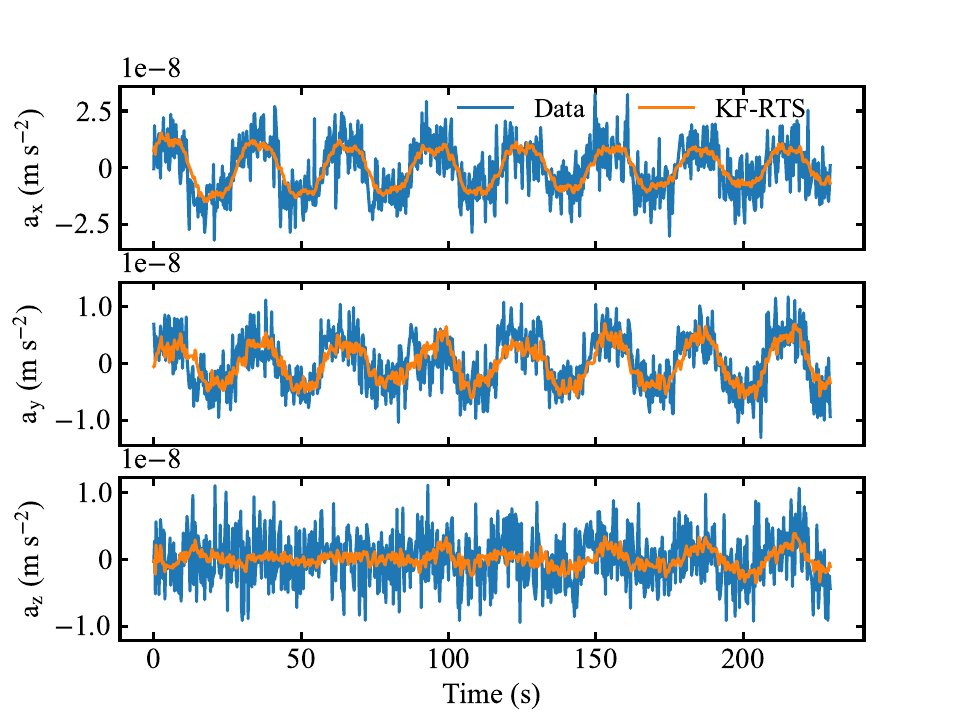}
\caption{The final round outliers detected through $\chi^2$ test and KF-RTS smoother result and calibration experiment data.}
\label{fig:Final_Outliers}
\end{figure}

Figure~\ref{fig:NLLS_without_outliers} illustrates the comparison between the fit results obtained using the NLLS algorithm and the experimental data with outliers removed. 
The final round results of the offset obtained using the KF-RTS algorithm are presented in Fig.~\ref{fig:Final_KF-RTS_dxdydz}, with the shaded area indicating the range of one standard deviation. 
Figure~\ref{fig:Final_Outliers} displays the comparison between the final round results obtained using the KF-RTS algorithm and the experimental data. 

Table~\ref{tab:result} presents the estimation results for the COM offset obtained using the two methods. 
Despite the use of different estimation algorithms, the results obtained from both methods are highly consistent, thus increasing the confidence in the accuracy of the results. 
It is worth mentioning that, the value of $\chi^2/\mathrm{nof}$ as the goodness of a fit is 2.33 for the first round and is 0.88 for the final round, where $\mathrm{nof}$ represents the number of degrees of freedom. 

\begin{table}
	\centering
	\caption{The results of estimation of the offset of COM.}
	\setlength{\tabcolsep}{1mm}{
			\begin{tabular}{cccc}
				\hline
				Axis & X ($\upmu$m) & Y ($\upmu$m) & Z ($\upmu$m) \\
				\hline
			    NLLS with outliers    & $-$175 $\pm$ 41 & 610  $\pm$  101 & $-$777 $\pm$  31 \\
                    First KF-RTS  & $-$173 $\pm$  27 & 606 $\pm$  66  & $-$777  $\pm$  20\\
				NLLS without outliers & $-$191 $\pm$  27 & 643  $\pm$  66  & $-$818  $\pm$  20 \\
				Final KF-RTS & $-$189 $\pm$  27 & 638  $\pm$  68  & $-$818  $\pm$  20\\
				\hline
	\end{tabular}}
	\label{tab:result}
\end{table}

\begin{figure}
\centering
\includegraphics[width=.48\textwidth]{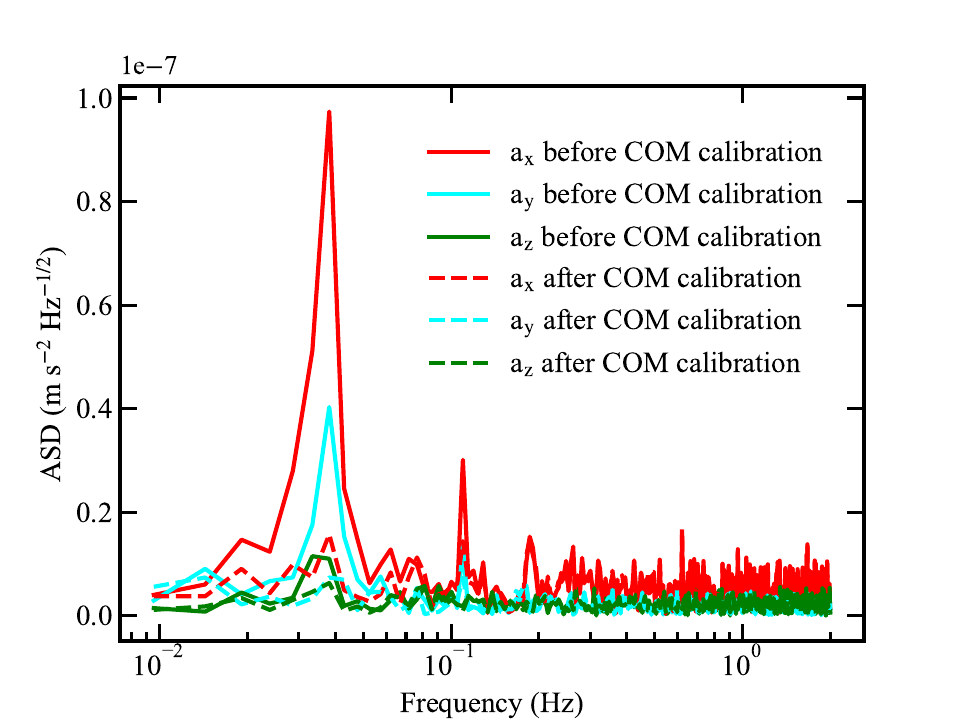}
\caption{The ASD of the calibration experiment data before and after the COM calibration.}
\label{fig:KF-TRS_Data_ASD}
\end{figure}

\begin{figure}
\centering
\includegraphics[width=.49\textwidth]{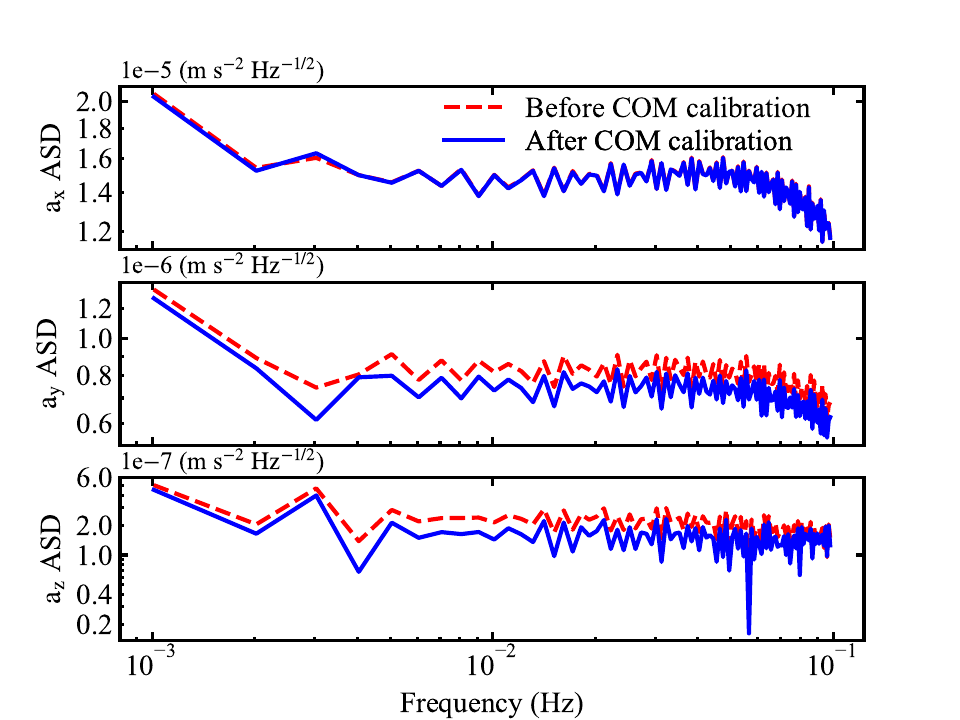}
\caption{The ASD of the flight data before and after the COM calibration.}
\label{fig:Real_Data_ASD_2022_3pic}
\end{figure}

Figure~\ref{fig:KF-TRS_Data_ASD} and Figure~\ref{fig:Real_Data_ASD_2022_3pic} depict the comparison of the amplitude spectral density (ASD) between the calibration experimental data and flight data before and after the COM calibration. 
Figure~\ref{fig:KF-TRS_Data_ASD} shows that the peak observed in the experimental data is caused by a modulation signal, and its influence is highly suppressed after the COM calibration, which is consistent with our expectations. 
Figure~\ref{fig:Real_Data_ASD_2022_3pic} demonstrates a significant reduction in the acceleration noise level of GRS readings in the frequency range of 0.001--0.1~Hz after the COM calibration. 

\section{Discussion and Conclusions}
\label{sec:Discussion and Conclusions}
Detecting and reducing the deviation between the COM of the inspection load and the COM of the satellite is crucial for high-precision accelerometers, as it can improve their accuracy. 
It is also a significant step in the development of space-based gravitational wave observatory to achieve their scientific objectives.
Furthermore, the calibration can help obtain a high-precision gravity field, which is valuable for conducting geoscience research with greater accuracy. 

In this study, the offset of the COM between the inspection load and the satellite is estimated using the KF-RTS smoother. 
Outliers are detected using the chi-square test, and the inverse Kalman filter is applied to remove them. 
The LM algorithm, as a cross check, is used to find the optimal offset parameters for the NLLS method. 
The results obtained with both methods are in very good agreement, and the offset of COM between the inspection load and the satellite is estimated with an accuracy of $\mathcal{O}(10\,\upmu\mathrm{m})$. 
After obtaining the COM offset, one can 
reduce it using the COM adjustment mechanism, and the effects of the COM offset can also be suppressed in the data-processing. 
The COM calibration is crucial for improving the accuracy of the accelerometer which will directly impact the detection sensitivity of the final space-based gravitational wave observatory. 

For the space-borne gravitational-wave observatories, such as Taiji-3, there are three satellites and each satellite is equipped with two TMs, which means that the COM of TMs does not coincide with that of the satellite. 
However, as a reference body for the satellite, 
it would cause residual acceleration 
if the TMs are away from their nominal positions. 
Therefore, it is necessary to periodically estimate or monitor the deviation of the COM of TM from a fixed point and make adjustments when necessary. 
The same calibration principle and methods as discussed in this paper can be used.
The Taiji-3 satellite will be equipped with higher-precision star-sensitive instruments, which will enable more accurate results. 

\section*{Acknowledgements}
This work is partially supported by the Strategic Priority Research Program of the Chinese Academy of Sciences Grant No. XDA15020700 and XDA15021100, and by the Fundamental Research Funds for the Central Universities. We acknowledge support from the National Space Science Data Center, National Science and Technology Infrastructure of China.

\bibliographystyle{apsrev4-2}
%

\end{document}